\documentclass[structabstract]{aa}

\usepackage{amsmath}
\usepackage{url}
\usepackage{txfonts}
\usepackage{natbib}
\usepackage{graphicx}
\bibpunct{(}{)}{;}{a}{}{,} 

\begin{document}

\title{Exploring the supernova remnant G308.4--1.4}
\author{T. Prinz \and W. Becker}

\institute{Max Planck Institute for extraterrestrial Physics, PO Box 1312, Giessenbachstr., 85741 Garching, Germany}

\date{Received <date> /
Accepted <date>}

\abstract {} {We present a detailed X-ray and radio wavelength study of G308.4--1.4, a candidate supernova remnant (SNR) in 
              the ROSAT All Sky Survey and the MOST supernova remnant catalogue, in order to identify it as a SNR.}
             {The SNR candidate and its central sources were studied using observations from the Chandra X-ray
             Observatory, SWIFT, the Australian Telescope Compact Array (ATCA) at 1.4 and 2.5 GHz and WISE infrared observation 
             at 24 $\mu$m.}
             {We conclude that G308.4--1.4 is indeed a supernova remnant by means of its morphology matching at X-ray, radio  
              and infrared wavelength, its spectral energy distribution in the X-ray band and its emission characteristics  
             in the radio band. G308.4--1.4 is a shell-type SNR. X-ray, radio and infrared emission is seen only in the eastern 
             part of the remnant. 
             The X-ray emission can best be described by an absorbed non-equilibrium collisional plasma with a hydrogen 
             density of $n_\mathrm{H}=(1.02\pm 0.04) \times 10^{22}$ cm$^{-2}$, a plasma temperature of $6.3^{+1.2}_{-0.7}$ million Kelvin 
             and an under-abundance of Iron, Neon and Magnesium, as well as an overabundance in Sulfur with respect to 
             the solar values. The SNR has a spectral index in the radio band of $\alpha=-0.7\pm0.2$ . 
	     A detailed analysis revealed that the remnant is at a distance of 6 to 12 kpc and 
             the progenitor star exploded $\sim 5000$ to 7500 years ago. Two faint X-ray point sources located near to the
             remnant's geometrical center are detected. Both sources have no counterpart at other wavelengths, leaving 
             them as candidates for the compact remnant of the supernova explosion.}
          {}


\maketitle

\section{Introduction}
 Supernovae (SNe), the core-collapse of massive stars or the thermonuclear disruption of white dwarfs, 
 are rare events, occurring on average every 50 years in the Galaxy \citep{2008MNRAS.391.2009K}. However,  
 only 7 Galactic SNe have been observed with historical records in the past two thousand years -- 
 SN 185 (RCW 86), SN 386 (G11.2--0.3), SN 1006, SN 1181 (3C58), Crab SN, Tycho SN and the Kepler SN. 
 Several other records of guest stars by ancient Chinese astronomers exists. However, no solid 
 identification with a known supernova remnant (SNR) has been possible for most of them so far 
 \citep{2003LNP...598....7G}. Most Galactic SNe go unobserved owing to visible-band extinction by 
 interstellar dust. For example SN Cas A may belong to this group. No widespread reports of Cas A exist  
 in the literature of the 17th century (cf.~Hartmann et al., 1997). A more recent example is the 
 youngest supernova known in our Galaxy, G1.9+0.3, which was totally missed by optical observatories 
 about 100 years ago \citep{2008ApJ...680L..41R}. The majority of supernova remnants 
 which have been identified so far were found in radio surveys in which the observations are 
 unhampered by interstellar dust \citep[and references therein]{2009BASI...37...45G}.
 Nevertheless, the radio band is not free of selection effects. SNRs with a diameter of $<3'$ are 
 difficult to identify. Only three of the 274 remnants listed in the Green catalogue fall into 
 this category. In addition, most searches for SNRs focus on the Galactic plane, leaving remnants 
 located at a higher galactic attitude underrepresented. Only 5 supernova remnants with $|b|>7^\circ$ 
 are listed in Greens catalogue. Hence, small remnants at high attitude are easily missed in 
 radio surveys.
 
 Most SNRs emit thermal radiation when the SN blast wave expands into the surrounding interstellar 
 medium (ISM), forms a shock wave at the shock front which then ionizes the atoms and increases the 
 temperature to $10^6 - 10^7$ K. Therefore, the successful completion of the ROSAT All-Sky Survey (RASS) 
 in the X-ray regime provided a new window for both finding SNRs and the compact objects 
 that may reside within them.  The seven X-ray detected and radio-quiet Central Compact Objects which 
 were associated with SNRs \citep[see][for a review]{2008AIPC..983..320G} are a good example for the 
 impact on pulsar and SN science of X-ray observatories. Additionally, X-ray observations of SNRs at 
 moderate spectral resolution can determine remnant properties such as shock velocities, post-shock gas 
 densities and temperature as well as swept-up shell mass and overall morphology.

 The ROSAT All-Sky survey has demonstrated the potential power for finding new SNRs 
 \citep{1996rftu.proc..267P, 1996rftu.proc..239B, 1994A&A...284..573A, 1996rftu.proc..233A, 
 1996rftu.proc..247E,  1996rftu.proc..253F}. This has motivated \citet{1998PhDTBusser} and 
 \citet{2002ASPC..271..391S} to investigate the all-sky survey data more systematically in order to
 search for unidentified supernova remnants. They analyzed RASS data in a systematic search 
 for extended ($\ge 5'$) X-ray sources at Galactic latitudes ($b \le \pm 15^\circ$) by correlating 
 them with databases like SIMBAD, NED, NVSS, SUMSS, DSSII, NRAO, ATNF, Parkes and Effelsberg 
 radio survey data. From their candidate list several sources have been confirmed meanwhile as SNR, e.g.
 \citep{2008ApJ...674..936J, 2012MNRAS.419.2623R, 2008ApJ...681..320R, 2007MNRAS.381..377S, 2010ApJ...712..790T}. 
 These identifications suggest that many of their SNR candidates are indeed radio-under-luminous, explaining 
 why past radio surveys could not detect or identify them before.

\begin{figure}
  \resizebox{\hsize}{!}{\includegraphics[angle=-90, clip]{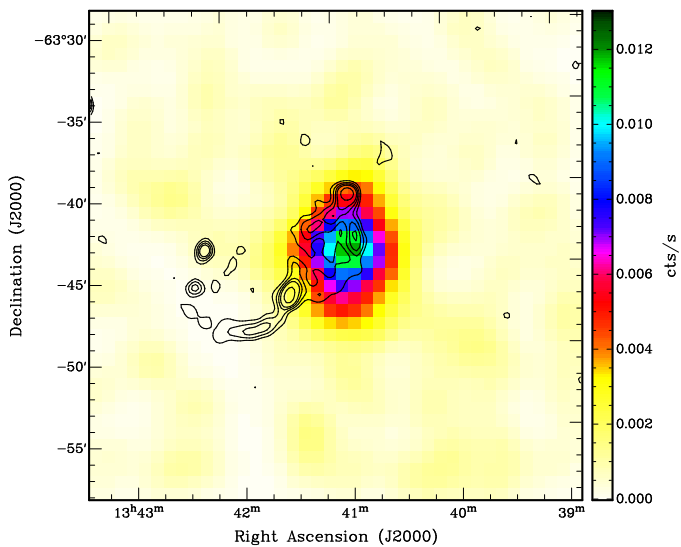}}
  \caption{ROSAT RASS $30'\times 30'$ image of G308.4--1.4 (0.1-2.4 keV) overlaid with the SUMSS image at 843 MHz 
  (contour lines at 4, 8, 12, 16 and 25 mJy). The half-power beam-width of the telescope is $43''$. 
  The RASS image is binned with $0.75'$ per pixel and smoothed by a Gaussian kernel of $\sigma \approx 2'$.}
  \label{fig.rass_most}
\end{figure}

 In \citet{2003PhDTSchaudel} 9 sources are listed which appear to be very 
 promising SNR candidates. One of them is G308.4--1.4 designated with G308.3--1.4 in 
 \citet{2002ASPC..271..391S,2003PhDTSchaudel}. Since the RASS had only an average exposure time of $\sim 400$s for sources in 
 the Galactic plane the available ROSAT data do not support a detailed spectral and spatial analysis of 
 these sources. However, G308.4--1.4 appears in the RASS data as a spherical center-filled X-ray source whereas 
 in the Sydney University Molonglo Sky Survey (SUMSS) two radio arcs have been detected which partly overlap 
 with the RASS data (see Fig. \ref{fig.rass_most}). Moreover, G308.4--1.4 is listed in the MOST supernovae 
 remnant catalogue as possible SNR candidate \citep{1996A&AS..118..329W}.

 In this paper we report on the identification of G308.4--1.4 as SNR using archival data from Chandra, SWIFT, 
 ATCA and WISE observatories. The results of the spatial and spectral analysis of the X-ray and radio data are 
 presented in \S \ref{kap:obs_x-ray} and \S \ref{kap:obs_radio}, respectively. In \S \ref{kap:obs_multi} 
 we report on data of G308.4--1.4 in other wavelength regimes. A concluding discussion about the 
 identification of G308.4--1.4 and its central sources is given in \S \ref{kap:discus_concl}. Furthermore, in the 
 same section we use the inferred spectral parameters of the SNR to derive an estimate for the age, the radius and the 
 expansion velocity as well as the distance to the remnant. Paragraph \ref{kap:Summary} provides a summary.

\section{X-ray observations and data analysis}\label{kap:obs_x-ray}

 The SNR candidate G308.4--1.4 was observed with the Advanced CCD Imaging Spectrometer (ACIS-I) on board the Chandra X-ray 
 observatory on 26th and 27th of June 2010 (obs.ID. 11249) for $\approx 15.1$ ks. All data were uncontaminated by
 soft proton flares. After dead-time correction we were left with 14.9 ks of good data. Standard processed  
 data were used with up-to-date calibration applied. The reduction of the data was done with 
 the Chandra Interactive Analysis software CIAO, version 4.3. For the spectral fitting we used the 
 X-Ray Spectral Fitting Package (XSPEC) developed by NASA, version 12.7.0. 

 In addition to Chandra we made use of the SWIFT observatory to analyze X-ray point sources detected in the central
 region of the remnant. G308.4--1.4 was observed three times with the Swift X-ray Telescope (XRT) on 17th of June 2011
 (obs.ID 00032030001), 7th (obs.ID 00032030002) and 11th of August 2011 (obs.ID 00032030003) for 3.5 ks, 4.2 ks 
 and 1.3 ks, respectively. These data were reduced with the HEASOFT software package, version 6.11. For the spatial 
 and spectral analysis we used the high resolution Chandra data as it contains five times more counts of G308.4--1.4. 

 To calculate the error in the X-ray flux we used a public available XSPEC 
 tool\footnote{\sffamily fluxerror.tcl\normalfont, see \url{http://heasarc.nasa.gov/xanadu/xspec/fluxerror.html}}.
 For the spectral analysis of the extended source emission we restricted the energy range to $0.5-4.0$ 
 keV as the count rate detected at higher energies is too sparse for a meaningful spectral analysis. 
 Below 0.5 keV the detector and telescope response is not well established. 
 In all other cases the energy range was restricted to 0.1--5 keV. All errors in this publication represent the 
 1$\sigma$ confidence range for one parameter of interest, unless stated otherwise.

\subsection{Spatial analysis}\label{kap:spatial}

\begin{figure}
  \resizebox{\hsize}{!}{\includegraphics[clip]{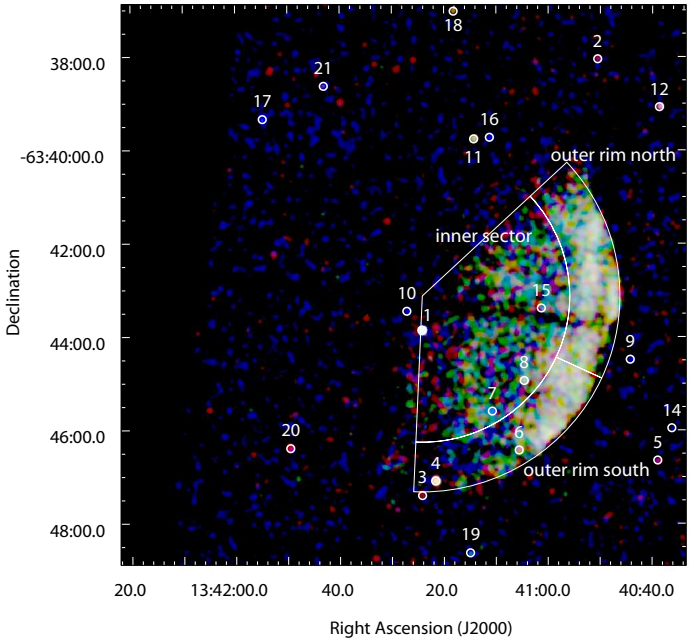}}
  \caption{Chandra ACIS-I color image of G308.4--1.4 (red 0.1-1.0 keV, green 1-1.5 keV and blue 1.5-5 keV). 
  The images are binned with $1''$ per pixel and smoothed by a Gaussian kernel of $\sigma = 3''$. 
  20 point sources were detected on the $12'\times 12'$ area around the center of G308.4--1.4 (c.f. Table \ref{tab:point_src}).}
  \label{fig.chandra_rgb}
\end{figure}

 Inspecting Fig.\ref{fig.chandra_rgb}, the X-ray image of G308.4--1.4 shows strong emission in the eastern part. This region has a circular shape with a diameter 
 of $\sim 8.4'$. The radiation is fading towards the geometrical center of the 
 source. The coordinate of the circle center is RA$_c = 13^\mathrm{h} 41^\mathrm{m} 23.9^\mathrm{s}\pm 0.7^\mathrm{s}$, 
 DEC$_c = -63^\circ 43' 08''\pm 04''$ which was obtained by fitting an annulus to the outer skirts of the extended 
 emission. 

 Searching for point sources in a $12'\times 12'$ box around G308.4--1.4 using a wavelet source detection algorithm 
 (CIAO-tool \sffamily wavdetect\normalfont) yielded a list of 21 sources.  Their position, positional error and signal-to-noise ratio 
 is listed in Table \ref{tab:point_src}. The sources itself are marked in Fig. \ref{fig.chandra_rgb}. Cross-correlating 
 the sources with the help of the VizieR online tool\footnote{\url{http://vizier.u-strasbg.fr/viz-bin/VizieR}} we found seven matches.
 They are included with their positional error in Table \ref{tab:point_src}.

 Eight of the detected sources are located within the extended emission of G308.4--1.4 and two of them (\# 1 and \# 10) are within 
 $\approx 1$ arcmin of the geometrical center of the SNR candidate.  Source 1 is the brightest 
 among all detected point sources and its designation is CXOU J134124.22--634352.0. Its position is coincident with 
 2MASS J13412422--6343520. In the optical source catalogs this source has no information about its proper motion. 
 Hence we cannot put better constraints on the positional coincidence of source \# 1 and the 2MASS object. 
 Source 10 has no optical counterpart and is designated with CXOU J134127.12--634327.7.

\begin{table*}
\begin{footnotesize}
\caption{Detected sources in a $12'\times 12'$ box around G308.4--1.4. The sources are denoted as in Fig. \ref{fig.chandra_rgb}.}
\label{tab:point_src}
\begin{tabular}{ccccccc}
\hline\hline \\[-2ex]
Source 	& RA (J2000) 	& DEC (J2000) 	& $\delta$ RA 	& $\delta$ DEC 	& $S/N$ & optical counterpart (positional discrepancy) \\ \hline
	& h:m:s		& d:m:s		& arcsec	& arcsec	& $\sigma_G$\tablefootmark{\alpha} & \\ \hline
1	& 13:41:24.217 & -63:43:52.04 & 0.41 & 0.40 & 63.0 & 2MASS J13412422--6343520 (0.07$'$)\\ 	
2	& 13:40:50.534 & -63:38:03.12 & 0.45 & 0.42 & 2.7 & \\			
3	& 13:41:24.203 & -63:47:24.47 & 0.70 & 0.42 & 3.1 & \\
4	& 13:41:21.634 & -63:47:05.63 & 0.44 & 0.43 & 25.5 & 2MASS J13412162--6347053 (0.27$'$)\\
5	& 13:40:38.921 & -63:46:38.83 & 0.54 & 0.45 & 3.3 & 2MASS 13403887--6346384 (0.48$'$)\\
6	& 13:41:05.585 & -63:46:25.89 & 0.47 & 0.42 & 4.9 &\\ 
7	& 13:41:10.742 & -63:45:36.11 & 0.52 & 0.42 & 4.9 & \\
8	& 13:41:04.580 & -63:44:56.66 & 0.52 & 0.42 & 6.7 & \\			
9	& 13:40:44.230 & -63:44:29.50 & 0.65 & 0.41 & 2.4 & \\
10	& 13:41:27.121 & -63:43:27.72 & 0.56 & 0.44 & 3.8 & \\
11	& 13:41:14.267 & -63:39:45.91 & 0.57 & 0.43 & 9.7 & 3UC 053--222814 (0.02$'$)\\	
12	& 13:40:38.666 & -63:39:04.77 & 0.61 & 0.43 & 9.0 & 2MASS J13403862--6339044 (0.41$'$)\\	
13	& 13:41:02.437 & -63:38:00.00 & 0.56 & 0.44 & 0.8 & \\
14	& 13:40:36.217 & -63:45:57.41 & 0.47 & 0.44 & 2.0 & \\
15	& 13:41:01.308 & -63:43:23.50 & 0.40 & 0.40 & 2.9 & \\
16	& 13:41:11.249 & -63:39:44.07 & 0.76 & 0.44 & 4.7 & \\
17	& 13:41:54.767 & -63:39:20.68 & 0.91 & 0.67 & 6.7 & 2MASS J13415473--6339207 (0.22$'$)\\	
18	& 13:41:18.147 & -63:37:01.87 & 1.31 & 0.53 & 3.8 & \\
19	& 13:41:14.934 & -63:48:38.00 & 0.67 & 0.54 & 3.2 & \\
20	& 13:41:49.497 & -63:46:23.59 & 1.06 & 0.65 & 4.1 & \\
21	& 13:41:43.058 & -63:38:38.10 & 1.24 & 0.50 & 5.4 & 2MASS J13414304--6338371 (0.93$'$)\\	
\hline\\[-1ex]
\end{tabular}
\tablefoot{
\tablefoottext{\alpha}{$\sigma_G=1+\sqrt{c_\mathrm{bg}+0.75}$, where $c_\mathrm{bg}$ are the background counts.}
}
\end{footnotesize}
\end{table*}

\subsection{Spectral analysis of G308.4--1.4}\label{kap:spec_whole_remnant}

To extract the spectra of the SNR candidate we first removed all point like 
sources and then selected all photons in a circular sector with a radius of 
$4.2'$ and opening angle of $\approx 136^\circ$ (see Fig. \ref{fig.chandra_rgb}). 
The background spectrum was derived from all photons in an annular sector with 
the same opening angle as the circular one but with the radii 5.5$'$ and 7.2$'$. 
After subtraction of the background contribution (0.054 cts/pix) we were left with 
11\,752 counts for the spectral analysis of the whole remnant-candidate. Additionally, the emission region of 
G308.4--1.4 was divided into four separate regions which are illustrated in Fig. \ref{fig.chandra_rgb}. 
The net counts are 4095 for the inner sector and 7658 for the outer shell, 
as well as 4499 for the northern part of the outer shell and 3158 
for the southern part. The spectrum of every single region was binned with at least 
50 counts per bin. Only the spectra of the southern and northern parts of the outer rim were 
binned with at least 75 counts per bin. 

To fit the X-ray spectrum of G308.4--1.4 we tried various one- and two-component spectral 
models. However, only a model of an absorbed collisional plasma which is in non-equilibrium and
allows a temperature evolution \citep{2001ApJ...548..820B} reproduces the energy distribution 
of the extended source. In XSPEC this model is implemented as \sffamily VGNEI\normalfont. 
To improve the goodness of the fit and to obtain the abundance of metals in the plasma we thawed 
the parameter of every single metal and monitored the improvement of the fit statistic.

\begin{figure}
\begin{center}
\includegraphics[width=0.375\textwidth,angle=-90,clip]{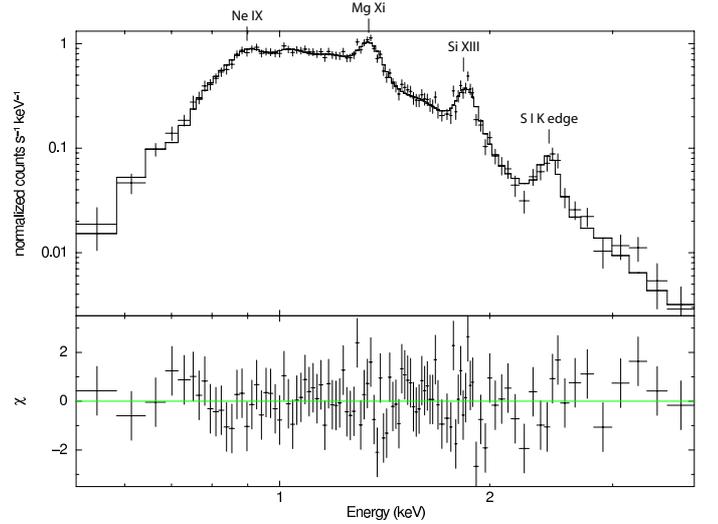}
\caption{Spectrum and fitted model of the X-ray emission of the whole remnant.}  
\label{fig:spec_remnant}
\end{center}
\end{figure}

The analysis results in a best fit model with a reduced $\chi^2$ of 1.03 for 95 degrees of freedom (d.o.f.) 
and a hydrogen column density $n_\mathrm{H}=1.02\pm 0.04\times 10^{22}$ cm$^{-2}$. The value for $n_\mathrm{H}$ is smaller than 
the integrated hydrogen column density towards G308.4--1.4 which is $1.31\times 10^{22}$ cm$^{-2}$ \citep{2005A&A...440..775K} 
and $1.45\times 10^{22}$ cm$^{-2}$ \citep{1990ARA&A..28..215D}. These values are based on HI emission line measurements at 21 cm and 
therefore refer to the entire hydrogen column density in the line of sight.
The temperature of the plasma is found to be $6.3_{-0.7}^{+1.2}$ million Kelvin. The best fit was obtained when thawing the abundance in 
Neon, Magnesium, Sulfur and Iron which differs significantly from the solar values: $\mathrm{Ne}=(0.72\pm 0.14)$ $\mathrm{Ne}_{\odot}$, 
$\mathrm{Mg}=0.70_{-0.08}^{+0.09}$ $\mathrm{Mg}_{\odot}$, $\mathrm{S}=2.3_{-0.8}^{+0.7}$ $\mathrm{S}_{\odot}$ and $\mathrm{Fe}=0.70_{-0.12}^{+0.14}$ 
$\mathrm{Fe}_{\odot}$. The ionization timescale $\tau_0=t_0 N_\mathrm{e}$ ($t_0$ is the remnant's age and $N_\mathrm{e}$ the post-shock
electron number density)  is $2.7\times 10^{10}$ s/cm$^3$ and the ionization timescale-averaged temperature 
$\langle kT \rangle=\int_{t_\mathrm{s}}^{t_0}T(t)N_\mathrm{e}(t)\mathrm{dt}/\tau=1.2_{-0.4}^{+2.2}$ keV ($t_\mathrm{s}$ is 
the age of the shock). Thus, the unabsorbed flux of G308.4--1.4 in the energy band $0.5 - 4.0$ keV is $f_X=1.61_{-0.23}^{+0.15}\times 10^{-10}$ 
ergs cm$^{-2}$s$^{-1}$. The model fit and the raw spectrum are shown in Fig. \ref{fig:spec_remnant}. The fitted 
parameters are listed in Table \ref{tab:spec_param}. 

Adding a second plasma component which is in non-equilibrium results in $\chi^2_{red}=1.06$ for 91 d.o.f. and hence 
is no improvement over a single component model. The other models we have fitted to the spectral distribution 
are a hot diffuse gas model (\sffamily VMEKAL\normalfont, $\chi^2_{red}=2.92$ for 96 d.o.f.), a model for a 
collisionally-ionized diffuse gas (\sffamily VAPEC\normalfont, $\chi^2_{red}=2.34$ for 96 d.o.f.) 
and a thermal bremsstrahlung model ($\chi^2_{red}=7.7$ for 101 d.o.f.). Non of these models can reproduce 
the spectrum of G308.4--1.4. However, all three model fits result in $n_\mathrm{H}\approx 1\times 10^{22}$ cm$^{-2}$ and a plasma 
temperature of $\sim 0.5$ keV.  

\subsubsection{The inner sector and the outer rim}\label{kap:part_remnant}

As mentioned in chapter \ref{kap:spec_whole_remnant} the extended source was divided into four regions. This was done 
to investigate the spatial variation of the hot plasma. For the extraction of
the background spectra of the inner sector and the outer rim we used the same area on the detector as for the whole remnant. 
The northern and southern region of the outer rim were analyzed separately. For the background we used only the part of this annular 
sector which has the same opening angle as the source region.

\begin{figure*}[!]
\begin{center}
\includegraphics[width=0.33\textwidth,angle=-90,clip]{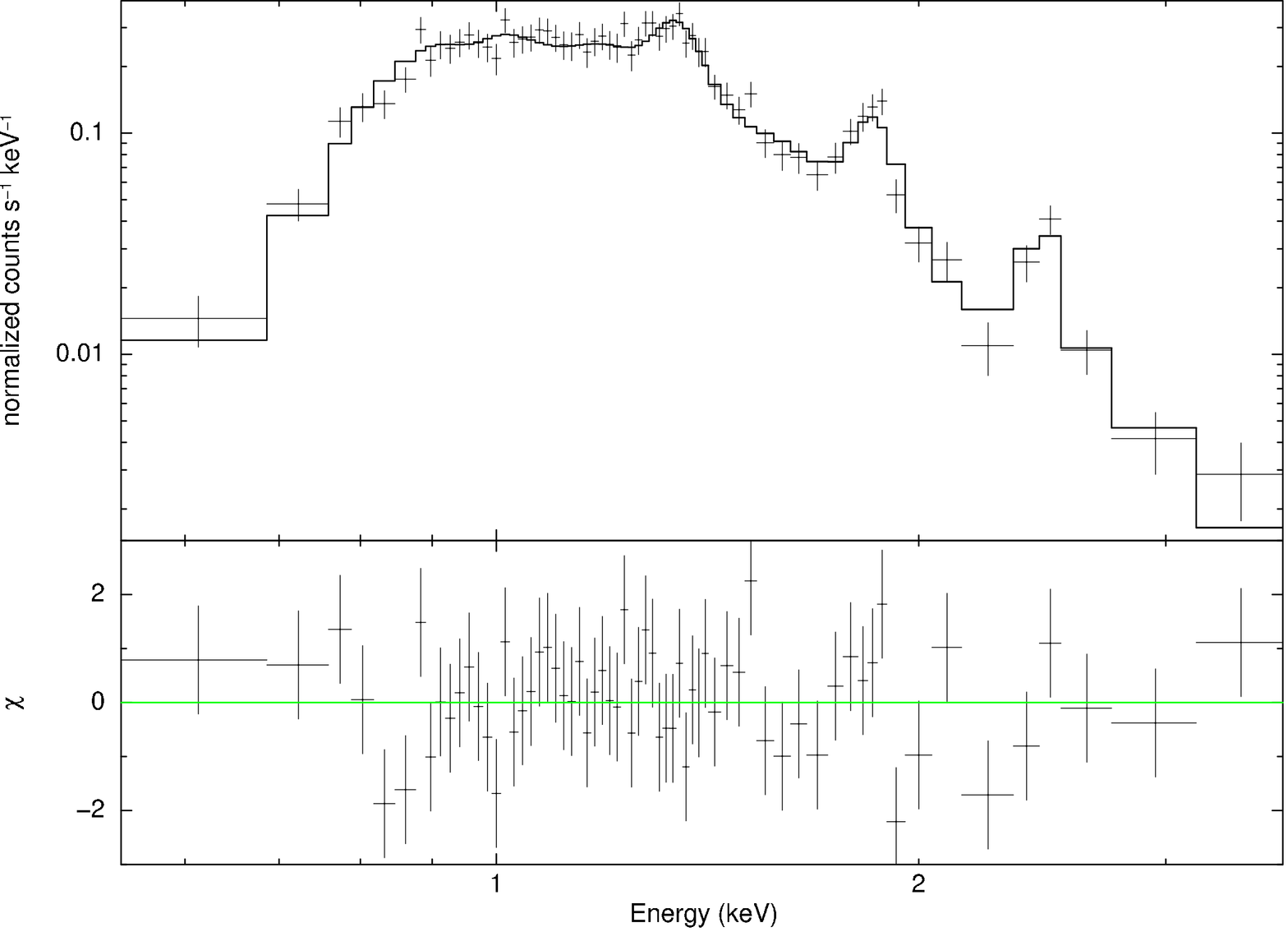}
\includegraphics[width=0.33\textwidth,angle=-90,clip]{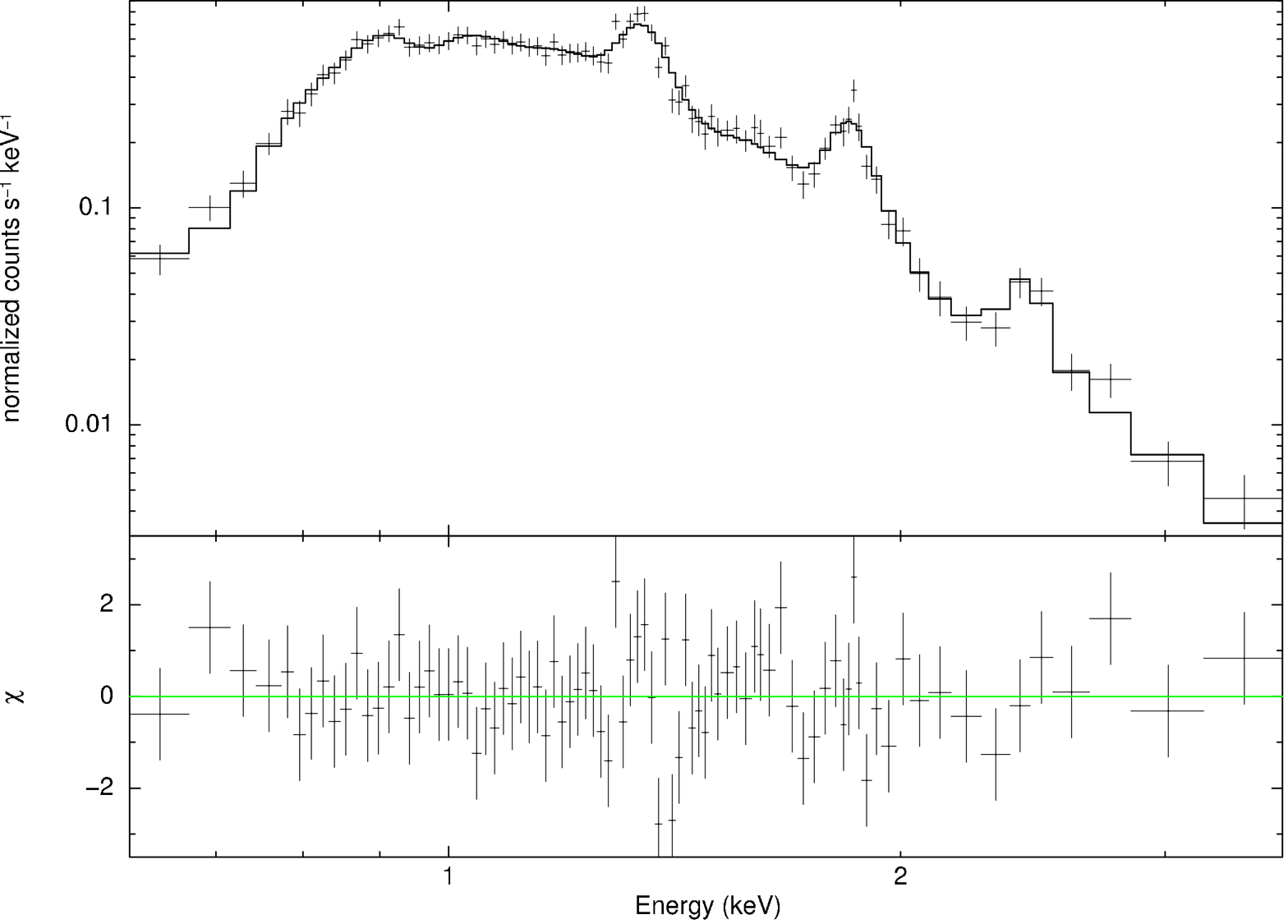}

\includegraphics[width=0.33\textwidth,angle=-90,clip]{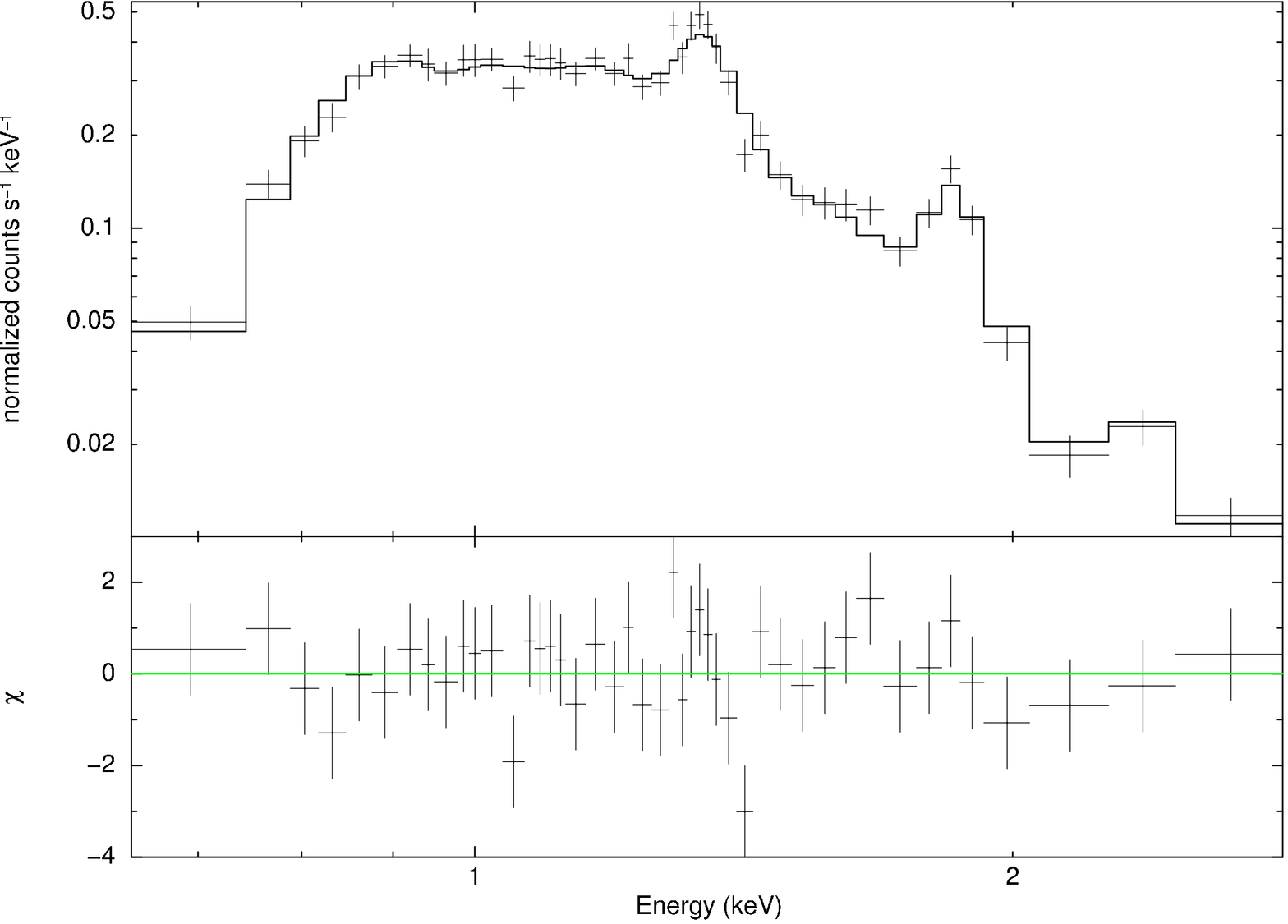}
\includegraphics[width=0.33\textwidth,angle=-90,clip]{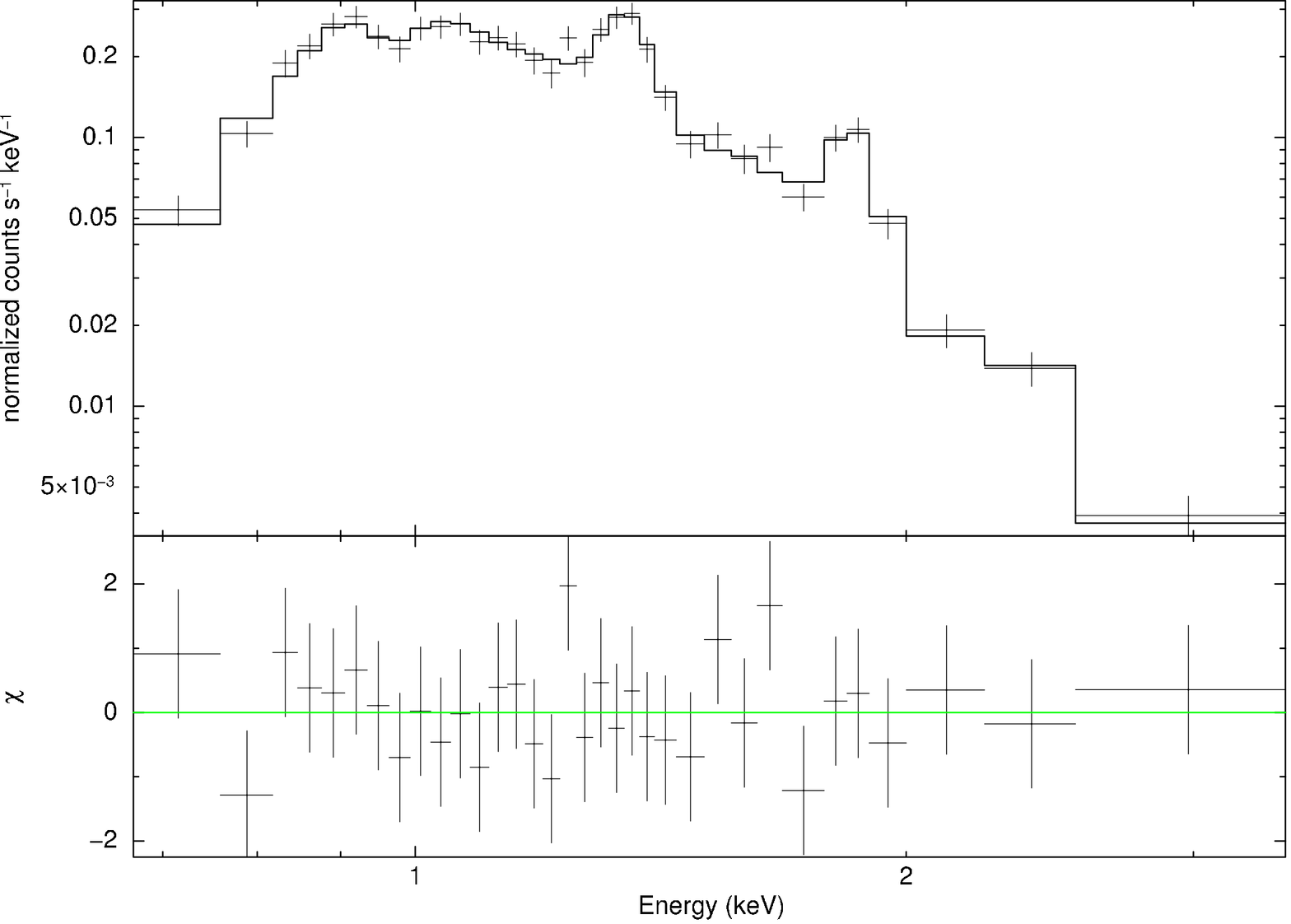}
\caption{Spectrum and fitted model of the X-ray emission of the various regions.
{\it Upper left}: inner part of the remnant, {\it upper right}: outer rim, {\it 
lower left}: outer rim north and {\it lower right}: outer rim south.}  
\label{fig:spec_regions}
\end{center}
\end{figure*}

First we fitted the plasma model with free $n_\mathrm{H}$. In every of the four spectral fits we found a hydrogen 
column density which is in good agreement with the $n_\mathrm{H}$ of the whole remnant. Second we fitted 
the spectra with a $n_\mathrm{H}$ fixed to the value of the whole remnant. The fitted value for $n_\mathrm{H}$ and 
the results of the fits with a fixed column density are summarized in Table \ref{tab:spec_param}.  
The spectra with the respective fits are shown in Fig. \ref{fig:spec_regions}. 

\begin{table*}
\renewcommand{\arraystretch}{1.25}
\begin{footnotesize}
\caption{Spectral parameters of the best fit model for different parts of SNR G308.4--1.4.}
\label{tab:spec_param}
\begin{tabular}{cccccc}
\hline\hline \\[-2ex]
				& Whole remnant		& Inner rim		& Outer rim		& Outer rim north 	& Outer rim south \\ \hline
$n_\mathrm{H}$ [$10^{22}$cm$^{-2}$]& $1.02\pm 0.04$	&$1.05_{-0.09}^{+0.11}$ & $1.00\pm 0.05$	& $1.02\pm 0.09$	& $0.99 \pm 0.09$\\
$k_\mathrm{B}T$ [keV]		& $0.54_{-0.06}^{+0.10}$&$0.55_{-0.10}^{+0.08}$	& $0.54_{-0.05}^{+0.08}$& $0.50_{-0.06}^{+0.12}$& $0.57\pm 0.08$\\
Ne\tablefootmark{\alpha} [Ne$_{\odot}$]		& $0.72\pm 0.14$	&$0.7\pm 0.3$		& $0.73_{-0.14}^{+0.15}$& $0.5\pm 0.2$		& $0.8\pm 0.2$\\
Mg\tablefootmark{\alpha} [Mg$_{\odot}$]		& $0.70_{-0.08}^{+0.09}$&$0.80_{-0.17}^{+0.20}$	& $0.67_{-0.09}^{+0.10}$& $0.83_{-0.13}^{+0.14}$& $0.62_{-0.12}^{+0.14}$\\
S\tablefootmark{\alpha} [S$_{\odot}$]			& $2.3_{-0.8}^{+0.7}$	&$3.0_{-1.0}^{+1.9}$	& $1.7_{-0.7}^{+0.8}$	& $2.4_{-1.0}^{+1.3}$	& $-$\\
Fe\tablefootmark{\alpha} [Fe$_{\odot}$]		& $0.70_{-0.12}^{+0.14}$&$0.9\pm 0.3$		& $0.64_{-0.12}^{+0.14}$& $-$			& $0.45_{-0.15}^{+0.19}$	\\
$\tau$\tablefootmark{\beta} [$10^{10}$ s/cm$^3$]	& 2.7			& 8.1			& 2.1			& 5.2			& 2.0\\
$\langle kT \rangle$\tablefootmark{\gamma} [keV]			& $1.2_{-0.4}^{+2.2}$	& $0.72_{-0.12}^{+0.75}$& $1.4_{-0.5}^{+1.6}$	& $0.81_{-0.15}^{+0.76}$& $1.1_{-0.4}^{+1.4}$\\
Norm\tablefootmark{\delta} [10$^{-2}$ cm$^{-5}$]	& $3.3\pm0.9$		& $0.9_{-0.2}^{+0.5}$	& $2.4\pm 0.5$		& $1.4\pm 0.4$		& $1.0\pm 0.3$\\
$\chi^2$			& 98.19			& 57.54 		& 76.81			& 38.74			& 18.55\\
d.o.f.				& 95			& 57			& 77			& 38			& 27\\
$f_X$\tablefootmark{\epsilon} [$10^{-10}$ ergs cm$^{-2}$s$^{-1}$]&$1.61_{-0.23}^{+0.15}$&$0.30_{-0.02}^{+0.01}$&$1.25_{-0.09}^{+0.08}$&$0.55_{-0.04}^{+0.02}$&$0.56_{-0.06}^{+0.07}$\\
\\[-2ex] \hline\\[-1ex]
\end{tabular}
\tablefoot{
\tablefoottext{\alpha}{abundance with respect to solar value}
\tablefoottext{\beta}{ionization timescale}
\tablefoottext{\gamma}{ionization timescale-averaged temperature}
\tablefoottext{\delta}{Norm=$\frac{10^{-14}}{4 \pi [D_\mathrm{A}(1+z)]^2} \int N_\mathrm{e} N_\mathrm{H} \mathrm{dV}$, where $D_\mathrm{A}$ is the angular diameter distance to 
the source in cm, $N_\mathrm{e}$ and $N_\mathrm{H}$ are the post-shock electron and hydrogen densities in cm$^{-3}$, respectively.}
\tablefoottext{\epsilon}{X-ray flux in the energy range 0.5 to 4.0 keV} 
}
\end{footnotesize}
\end{table*}

Within the statistical uncertainties, all regions turned out to have
a similar temperature. Only the ionization timescale-averaged temperature was varying.
The large statistical uncertainty also prevents us from concluding on a
temperature evolution with the distance to the expected center of the remnant. 
No variation of the Neon and Magnesium abundance over the surface of the remnant is 
detectable. The only noticeable change 
in abundance is in Sulfur and Iron in the outer rim. The spectrum in the northern rim shows 
an abundance in Iron comparable with the solar value. In the southern rim no significant 
deviation from the solar value was found in the abundance of Sulfur. Furthermore, the ionization 
timescale varies with distance from the remnant's center and the direction of the expanding 
shock wave into the ISM. The highest values for $\tau$ were found for the inner sector 
($8.1 \times 10^{10}$ s/cm$^3$) and the lower one in the southern part of the rim 
($2.0 \times 10^{10}$ s/cm$^3$) which implies a different post-shock electron number 
density in the inner and outer regions of the SNR candidate. 
 
\subsection{The central sources}

As mentioned above, two point sources were detected close to the presumed center of the remnant, 
source \# 1 (CXOU J134124.22--634352.0) and \# 10 (CXOU J134127.12--634327.7). For the spectral analysis of source \# 1 we used all 
photons within a circular region of radius 10 arcsec centered on the source position. The background 
was extracted from an annulus with radii $15''$ and $20''$. After background correction we were left 
with 195 source counts in the energy band between 0.5 and 10.0 keV. We used the CIAO tool \sffamily psextract 
\normalfont to compute the response matrices which are necessary to compensate for the changing spectral response
of the CCD and the telescope with energy and off-axis angle. The spectrum was binned with at least 20 counts per bin. 

\begin{figure}
  \resizebox{\hsize}{!}{\includegraphics[clip]{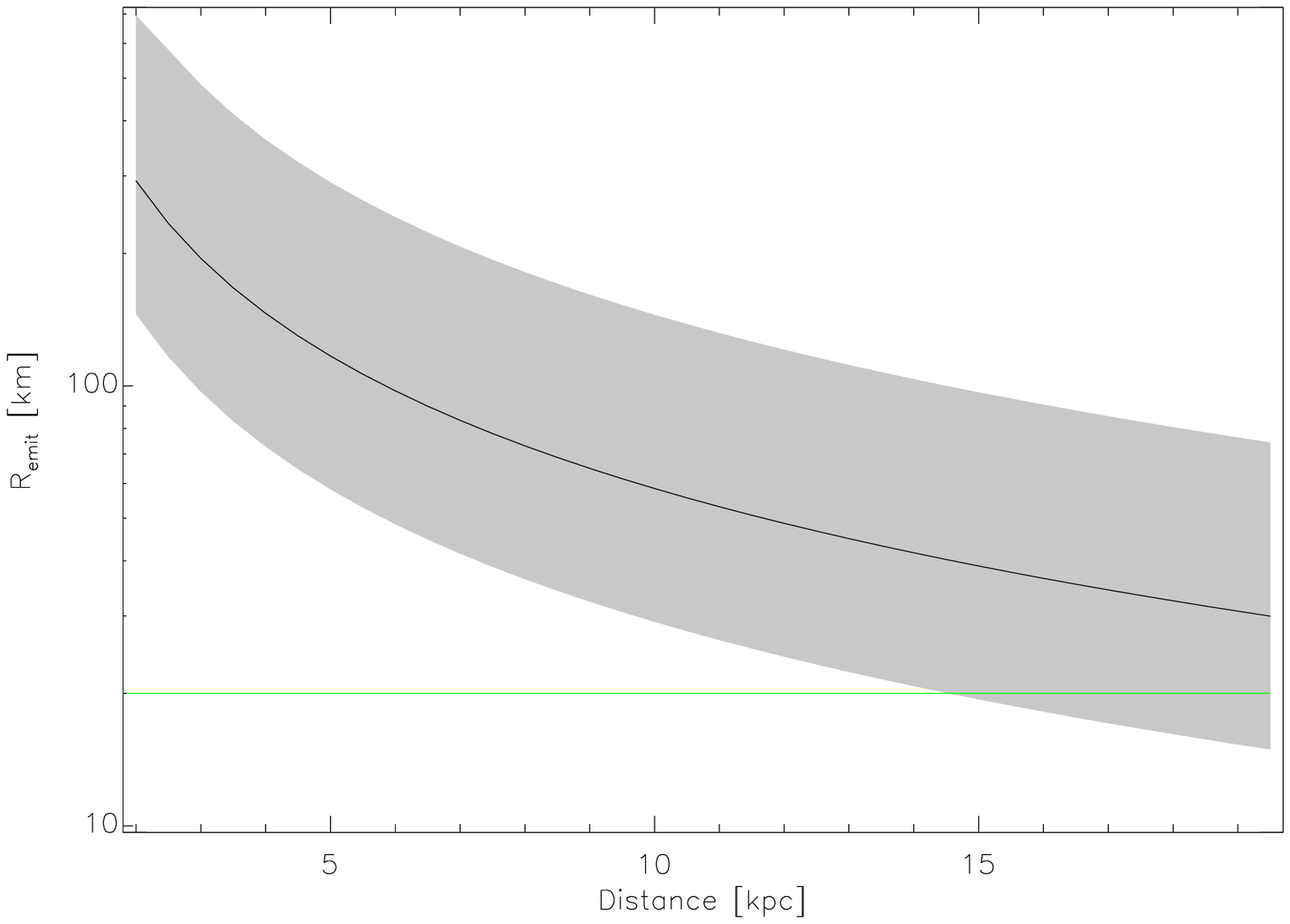}}
  \caption{Dependence of the blackbody emitting radius $R_1$ of the first blackbody in the double blackbody spectral fit 
on the distance to source \# 1 in kpc (equation \ref{equ.Remit_vs_D}). The error range is gray shaded and
the green line indicates the upper limit on the radius of a neutron star.}
  \label{fig.R_vs_d}
\end{figure}

\begin{figure}
  \resizebox{\hsize}{!}{\includegraphics[clip,angle=-90]{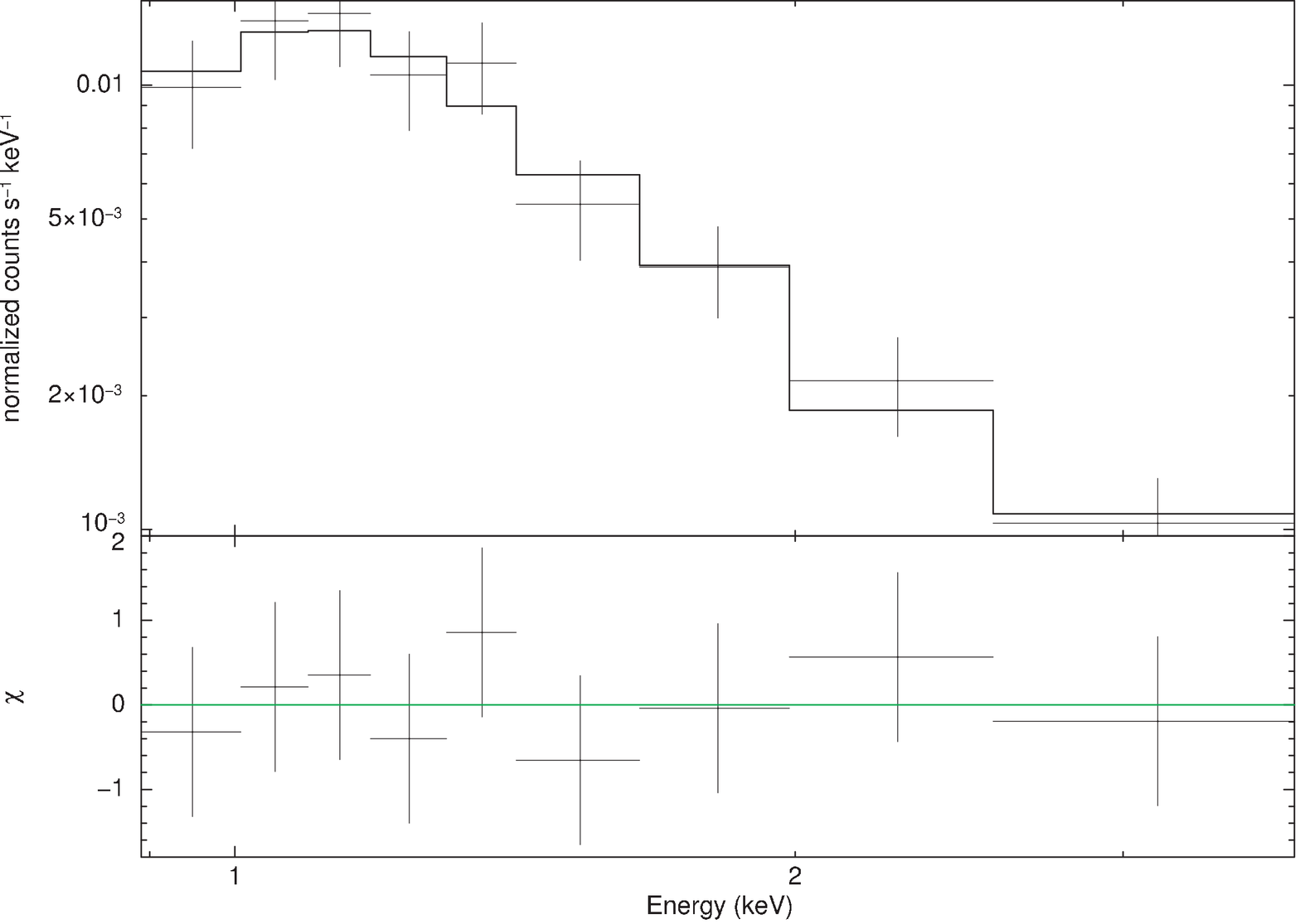}}
  \caption{Spectrum of source \# 1 fitted with a double blackbody spectrum.}
  \label{fig.spec_src1}
\end{figure}

A single absorbed blackbody or a single absorbed power law model for fitting the spectral distribution 
gives poor results. In the case of a blackbody spectrum the reduced $\chi^2=2.67$ for 6 degrees of 
freedom is unacceptably large. The power law model fit results in a reduced $\chi^2$ of 0.94 
(6 d.o.f.). However, the post-fit residuals show systematic derivations between the data points 
and the fitted model. Therefore, we used a two-component absorbed blackbody model which results in 
$\chi^2= 1.62$ (4 d.o.f.) and a $n_\mathrm{H}=1.34^{+1.80}_{-1.31} \times 10^{22}$ cm$^{-2}$. To better 
constrain the spectral parameters we fixed $n_\mathrm{H}$ to the hydrogen column density 
obtained from fits to G308.4--1.4. This gave a reasonable fit with $\chi^2=1.96$ for 5 d.o.f. (see Fig. 
\ref{fig.spec_src1}). The temperatures of the two blackbodies are $T_1=1.3_{-0.2}^{+0.3}\times 10^6$ K 
and $T_2=6.4_{-1.4}^{+2.8}\times 10^6$ K. The normalization factors of the two blackbodies are $K_1= 7_{-5}^{+21}\times 10^{-5}$ 
and $K_2= (1.3\pm +0.4)\times 10^{-6}$. Using these values and the distance to the source $D_{10}$ in 
units of 10 kpc we estimated the emitting source radius $R$ as follows:

\begin{equation}\label{equ.Remit_vs_D}
R_i=\sqrt{\frac{K_i \times 10^{39}}{4 \pi \sigma_B D^2_{10} T^4}}~\mathrm{[m]},
\end{equation} where $\sigma_B$ is the Stefan-Boltzmann constant. As we do not know the distance to 
G308.4--1.4 we considered the emitting radius as function of the distance to the source. This is illustrated 
in Fig. \ref{fig.R_vs_d}. The result for the second blackbody is $R_2=340_{-158}^{+302} D^{-1}_{10}$ m. 
The unabsorbed flux in the energy band 0.5 to 10 keV is $f_X=2.2_{-1.9}^{+0.5}\times 10^{-12}$ 
ergs cm$^{-2}$s$^{-1}$ and in the energy band 0.3 to 3.5 $f_X=(4\pm 2)\times 10^{-12}$ ergs cm$^{-2}$s$^{-1}$.

Additionally, we tried a model composed of a blackbody and a power law with the same fixed $n_\mathrm{H}$ as before. 
This gives a reasonable fit with $\chi^2=2.17$ (5 d.o.f.). The photon index is $2.4_{-1.2}^{+1.0}$ and the blackbody 
temperature is $(1.3\pm 0.3)\times 10^6$ K. With the normalization of $K_1=9_{-7}^{+53}\times 10^{-5}$ the  
emitting radius is $73_{-45}^{+216} D^{-1}_{10}$ km. 
The unabsorbed flux in the energy band 0.5 to 10 keV is $f_X=2.5_{-2.3}^{+0.5}\times 10^{-12}$ ergs cm$^{-2}$s$^{-1}$.

As mentioned in section \ref{kap:spatial} the position of source \# 1 is consistent with 2MASS J13412422--6343520, 
a star with a magnitude of $B=19.7$, $R=19.6$, $J=14.0$, $H=13.3$ and $K=13.1$. This implies an X-ray-to-visual 
flux ratio of $\log(f_X/f_V)=\log(f_X) + V/2.5 + 5.37=1.55$ \citep{1988Maccacaro}. Here $f_X$ is the X-ray flux in 
the $0.3-3.5$ keV band in ergs cm$^{-2}$ s$^{-1}$, $V$ is the visual magnitude and we assumed that $V\approx R$.

In addition we tried to fit various plasma models to the source spectrum. The only reasonable fit
was found when using a collisionally-ionized diffuse gas model (\sffamily APEC \normalfont). 
The goodness of fit which we found for 6 d.o.f. was $\chi^2_{red}=0.78$
for the following spectral parameters: $n_\mathrm{H} < 0.47 \times 10^{22}$ cm$^{-2}$, $ T=3.1^{+1.7}_{-1.4} \times 10^7$ K. 
For this model the flux in the 0.3 to 3.5 keV range is $f_X=1.1^{+0.2}_{-0.1}\times 10^{13}$ ergs cm$^{-2}$s$^{-1}$. 
Thus, the X-ray-to-visual flux ratio is $\log(f_X/f_V)=0.25$.
Again no variation in the spectral parameters between the flaring and quiescence epoch are detectable.
All fitted models and their inferred spectral parameters are listed in Table \ref{tab:point_src_spec}.

\begin{table*}
\linespread {1.25}\selectfont
\begin{footnotesize}
\caption{Spectral fit results of the central source \# 1.}
\label{tab:point_src_spec}
\begin{tabular}{cccccccc}
\hline\hline \\[-2ex]
Model			& $\chi^2$/d.o.f.& $n_\mathrm{H}$		& $\Gamma$/$kT_1$ 	& $K_1$ 				& 
$kT_2$ 		& $K_2$ 				& $f_X^{0.5-10}$/$f_X^{0.3-3.5}$ \\ 
			&		& $10^{22}$ cm$^{-2}$	& -/keV			& 	& keV		& 	& $10^{-12}$ergs cm$^{-2}$ s$^{-1}$ 	\\ \hline
BB\tablefootmark{\alpha,\delta}	& 16.00/6	& $1.2\times10^{-4}$	& $0.44$& $9.9\times10^{-7}$	& 
--&--& $-$/$-$\\
PL\tablefootmark{\alpha,\gamma}	& 5.62/6	& $<0.33$		& $2.2_{-0.4}^{+0.9}$	& $3.9_{-0.9}^{+4.3}\times 10^{-5}$	& 
--&--& $0.16_{-0.02}^{+0.03}$/$0.16_{-0.03}^{+0.04}$\\
BB+BB\tablefootmark{\beta}	& 1.96/5	& $1.34_{-1.31}^{+1.80}$& $0.11_{-0.02}^{+0.03}$	& $7_{-5}^{+21}\times 10^{-5}$	& 
$0.55_{-0.12}^{+0.24}$&$(1.3 \pm -0.4) \times 10^{-6}$& $2.2_{-1.9}^{+0.5}$/$4\pm 2$ \\
PL+BB\tablefootmark{\beta,\gamma}	& 2.74/5	& $<3.8$		& $2.4_{-1.2}^{+1.0}$	& $7_{-5}^{+11}\times 10^{-5}$	& 
$0.11\pm 0.03$&$9_{-7}^{+53}\times 10^{-5}$& $2.5_{-2.3}^{+0.5}$/$5_{-4}^{+3}$ \\
APEC		& 4.66/6	& $<0.47$	& $2.7_{-1.2}^{+1.5}$				& $9.0_{-1.5}^{+3.0}\times 10^{-5}$	& 
--&--& $0.13_{-0.02}^{+0.03}$/$0.11_{-0.03}^{+0.04}$\\
BREMS\tablefootmark{\alpha,\delta}	& 6.98/6& $1.8\times10^{-9}$& $2.1_{-0.8}^{+2.1}$		& $5.1_{-1.3}^{+2.3}\times 10^{-5}$	& 
--&--& $-$/$-$\\
RAYMOND\tablefootmark{\alpha}	& 5.70/6	& $<0.21$	& $2.8_{-1.1}^{+2.0}$			& $8.9_{-0.8}^{+1.8}\times 10^{-5}$	& 
--&--& $0.13\pm-0.02$/$0.11_{-0.01}^{+0.02}$\\
\\[-2ex]\hline \\[-1ex]
\end{tabular}
\linespread {1.0}\selectfont
\tablefoot{
\tablefoottext{\alpha}{fit shows systematic derivations between the data points and the fitted model.}
\tablefoottext{\beta}{the spectral parameters are calculated with $n_\mathrm{H}$ fixed to the value obtained in the fit to the spectrum of G308.4--1.4.}
\tablefoottext{\gamma}{Power law norm in photons keV$^{-1}$cm$^{-2}$s$^{-1}$ at 1 keV.}
\tablefoottext{\delta}{some errors could not be calculated.}
}
\end{footnotesize}
\end{table*}

\subsubsection{Temporal analysis of source \# 1}

\begin{figure}
  \resizebox{\hsize}{!}{\includegraphics[clip]{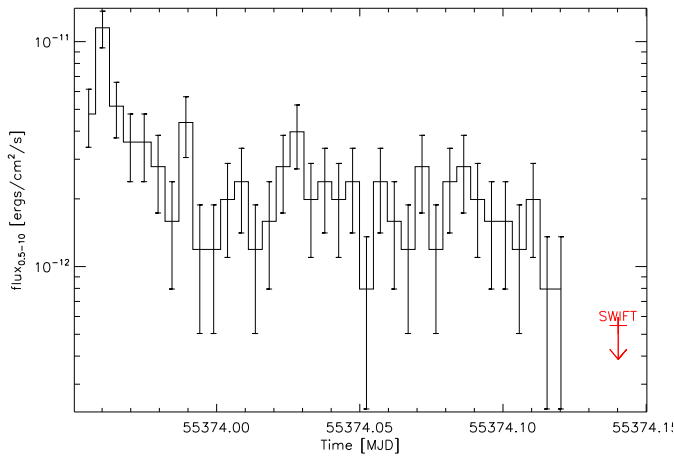}}
  \caption{Flux variation of source \# 1 in the Chandra observation binned with
  400 s per bin and scaled with the mean flux in the whole exposure. 
The red arrow indicates the 3$\sigma$ upper limit derived with the SWIFT observations.}
  \label{fig.flaring_src1}
\end{figure}

In order to search for any temporal variability we binned the photon time of arrivals with 
400 s per bin (see Fig. \ref{fig.flaring_src1}). The data show clearly a flare in the 
first kilo-seconds of the observation. The spectral parameter of the double blackbody in the flaring 
and quiescence epoch match the overall spectrum within the error bars. Varying only the 
normalizations of the two blackbodies the flux in the quiescence epoch is with $f_X^{0.5-10}= 
(1.9\pm 0.3)\times 10^{-12}$ ergs cm$^{-2}$s$^{-1}$ slightly lower than the flux in the total 
exposure. However, the flux in the time interval when the source was flaring is 
$f_X^{0.5-10}= (5.2 \pm 1.5)\times 10^{-12}$ ergs cm$^{-2}$s$^{-1}$, three times the 
quiescence flux. However, the sparse photon statistics lead to large uncertainties in the 
derived errors of the quiescence and flaring fluxes and thus they are only rough estimates.
Merging the SWIFT observations we cannot detect a source at the position of source \# 1. 
In the merged image a total of 4 counts were recorded within a circle with radius 15 arcsec 
(equals an encircled energy of 70~\%) centered on the position of source \# 1 . The
total exposure time of the source was 8344 s. Hence, the $3\sigma$ upper limit for the 
counting rate is approximately $8.6\times 10^{-4}$ cts/s. Using the mission count 
rate simulator for X-ray observatories WebPIMMS\footnote{\url{http://heasarc.nasa.gov/Tools/w3pimms.html}} with the SWIFT 
response and the spectral parameters derived for a double blackbody spectrum the upper limit 
on the flux is  $\approx 5.3\times 10^{-13}$ ergs cm$^{-2}$ s$^{-1}$. For the \sffamily APEC \normalfont model the 
upper limit is $\approx 2.3\times 10^{-14}$ ergs cm$^{-2}$ s$^{-1}$. These limits are less 
than 25 \% of the average flux in the Chandra observation, clearly indicating that the 
source flux is variable on time scales of at least several hours to weeks.
 
\subsubsection{Extent of source \# 1}

With the help of a one-dimensional distribution of the counts per $0.5''$ bin along a 
vertical line in the Chandra image around source \# 1 we checked if the source 
is extended. The resulting diagram was obtained by integrating all counts within 
$5''\times 0.5''$ rectangular apertures moving along the western direction with 
$0.5''$ steps. We compared this with the count distribution of a point source 
simulated with MARX\footnote{MARX is a software which can be used to simulate the 
performance of the Chandra X-ray observatory; see \url{http://space.mit.edu/cxc/marx/}}. 
However, no excess emission of point source \# 1 could be detected.

For source \# 10, the other source located close to the center of the remnant, 
we detected only 8 net counts. Hence, the count rate is $(5.1\pm 1.9)\times 10^{-4}$ 
cts/s, but no spectral analysis is possible by the limited statistics. 

\section{Radio observation and data analysis}\label{kap:obs_radio}
The source G308.4--1.4 was observed for 11.25 hours with the Australia Telescope 
Compact Array, a synthesis telescope near Narrabri, New South Wales on 11th
January 2002 at 1.432 and 2.448 GHz with 13 channels per band. 
The single pointing continuum observations of the source G308.4--1.4 was 
carried out with the array configuration 0.75C (maximal baseline length is 750 m). 
All four Stokes parameters have been recorded.

The flux density calibration was performed through observations of 
PKS B1934--638, which is the standard primary calibrator for ATCA observations. 
Phases were calibrated using observations of secondary calibrating sources 
PKS 1329--665. Since the primary beam response is frequency dependent we did not 
merge the data from two observing bands before imaging and cleaning. Every single 
observing run and each of the two observing bands were calibrated separately 
following the standard procedures for ATCA observations.

The reduction was carried out using the ATNF release of the Multichannel 
Image Reconstruction, Image Analysis and Display \citep[MIRIAD,][]{1995ASPC...77..433S}. 
A number of steps in the reduction process (i.e. ``flagging``) were done interactively. 

   Because of the bad phase stability during the observation, the scalar 
averaging of gains over the interval of 5 min (the length of secondary
calibrator's scan) was performed. Then, a total-intensity continuum image square 
region was formed using multi-frequency synthesis, uniform weighting and a cell 
size of 10 arcsec. At the stage of the dirty map production all the correlations 
with 6th baseline were excluded to obtain images with better signal-to-noise
ratio. The next step was to de-convolved both images using the standard CLEAN algorithm 
\citep{1980A&A....89..377C} with 10\,000 iterations. The resulting images were restored and 
corrected for the mean primary beam response of the ATCA antennas.
The final images have synthesized beam sizes of $63.6''\times 53.2''$ and $35.2''\times 29.3''$ 
for the 1.384 GHz and 2.496 GHz band, respectively.

\begin{figure*}[!]
\begin{center}
\resizebox{0.8\hsize}{!}{\includegraphics[clip]{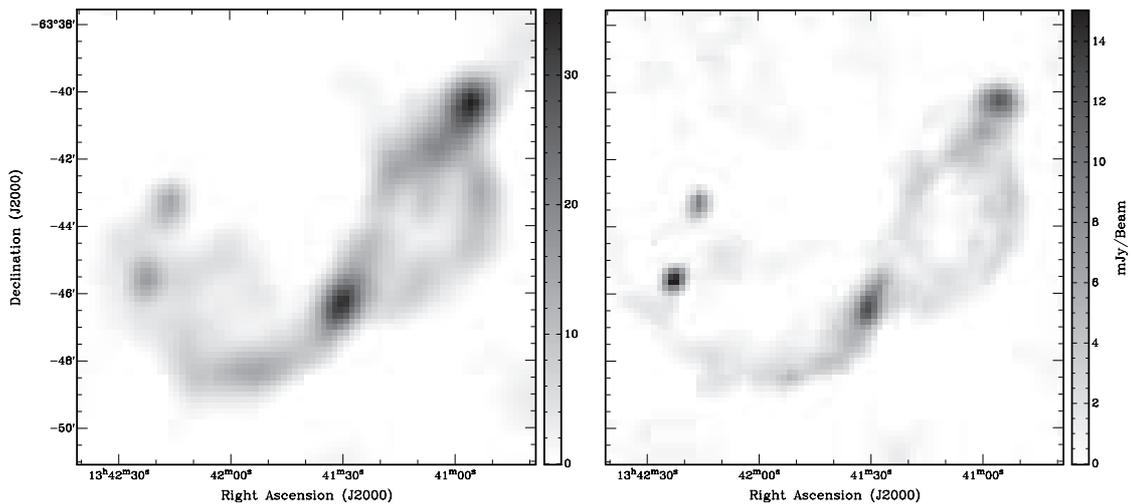}}
\caption{Grey-scale image of SNR G308.4--1.4. The left image displays the SNR at 1.4 GHz and on the right the same 
region at 2.5 GHz. The 1.4 GHz image is smoothed with a beam of $63.6''\times 53.2''$ and the 2.5 GHz with a beam of $35.2''\times 29.3''$.}  
\label{fig:becker}
\end{center}
\end{figure*}

\begin{figure}
  \resizebox{\hsize}{!}{\includegraphics[angle=-90, clip]{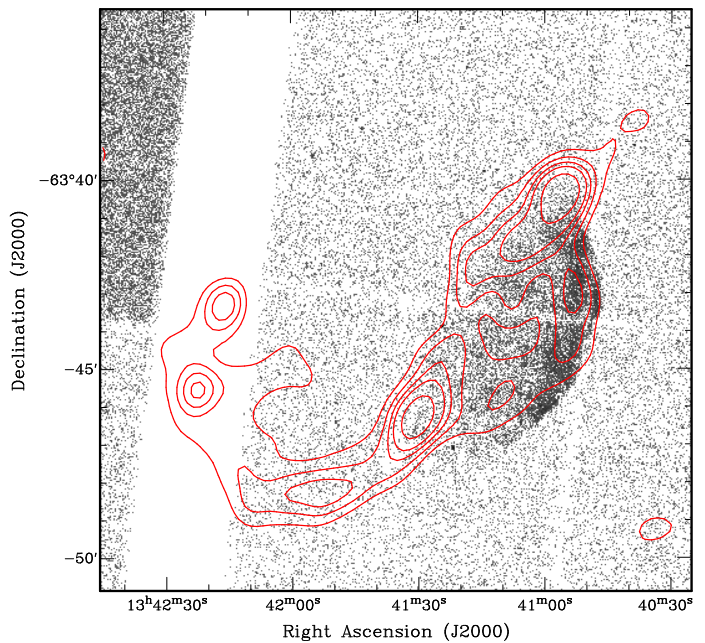}}
  \caption{Chandra $15'\times 15'$ image of G308.4--1.4 overlaid with the ATCA data at 1.4 GHz. 
  The contour levels are 4, 8, 12, 16 and 24 mJy beam$^{-1}$. The radio image has a full width at half maximum (FWHM) 
  of $ 63.6''\times 53.2''$ and the X-ray image is binned with $1''$ per pixel.} \label{fig.chandra_radio}
\end{figure}

In both frequency bands the images show two radio arcs (see Fig. \ref{fig:becker}). 
The eastern arc matches the X-ray emission region. For the western arc no counterpart is visible in the X-ray regime.
There are also two radio bright knots within the arcs which do neither have a counterpart in 
X-rays (see Fig. \ref{fig.chandra_radio}) nor in the optical band (dss2red). No difference 
can be seen to the morphology of the extended source in the SUMSS 843 MHz map. 

The flux density of G308.4--1.4 without the two bright knots are 235, 174 and 110 mJy at 
843, 1384 and 2496 MHz, respectively. These values were 
obtained by integrating the emission within multiple polygons enclosing the remnant. The 843 MHz flux 
density was deduced from the SUMSS image. The bright southern knot has a flux density of 110, 
75 and 70 mJy at 843, 1384 and 2496 MHz. Whereas for the northern knot we measured a flux density of 96, 76 and 56 mJy 
at 843, 1384 and 2496 MHz. In all cases we assumed a conservative error in the flux density of $20\%$.

The rms noise levels are 0.28 mJy beam$^{-1}$ and 
0.05 mJy beam$^{-1}$ for the 1.384 GHz and 2.496 GHz image. The levels were measured by integrating over 
nearby source-free regions. The spectral indices were computed using the total flux density measurements
in all available frequencies (0.843, 1.384 and 2.496 GHz). Fitting the data with a
power law $S_\nu \propto \nu^\alpha$, we obtained a spectral index for the northern knot 
$\alpha=-0.50\pm0.14$ and for the southern $\alpha=-0.49\pm0.16$, respectively. The radio 
emission of the two arcs, however, have a spectral index $\alpha=-0.7\pm0.2$. 
No significant polarized radio emission could be detected.

\section{Observations in other wavelength regimes}\label{kap:obs_multi}
\begin{figure}
  \resizebox{\hsize}{!}{\includegraphics[angle=-90,clip]{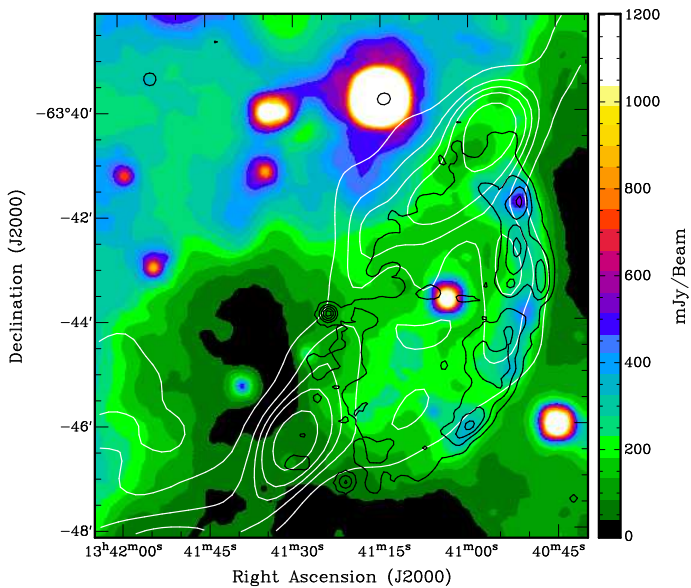}}
  \caption{$10'\times 10'$ WISE image around G308.4--1.4 at 22 micron with the radio contour lines in white 
at 3, 8, 12, 16 and 24 mJy and the X-ray contours in black.}
  \label{fig:infrared}
\end{figure}

The region around G308.4--1.4 was observed in the four infrared bands 3, 5, 12 and 22 $\mu$m during the 
Wide-field Infrared Survey \citep[WISE,][]{2010AJ....140.1868W}. With the 22 $\mu$m filter an extended 
infrared emission, overlapping with the radio and X-ray emission regions, is detectable (see Fig. \ref{fig:infrared}).
After subtraction of the background contribution the intensity image in the 22 $\mu$m band was converted into a flux 
image using the explanatory Supplement to the WISE All-Sky Data Release 
Products\footnote{\url{http://wise2.ipac.caltech.edu/docs/release/allsky/expsup/sec2_3f.html}}. We did not apply color correction as the 
emission mechanism in the infrared regime is uncertain. In addition, the integrated flux is not corrected for extinction.
To compare the infrared with the 1.4 GHz image the flux density in the 22 $\mu$m observation was scaled to Jy/beam, where beam is $63.6 \times 53.2$ arcsec$^2$. 
Using the region of infrared emission which is clearly associated with the remnant the integrated flux of G308.4--1.4 in the 22 $\mu$m band 
is $\approx 1.7$ Jy.

The shock wave of a SN which is running into the interstellar medium can interact with adjacent 
molecular clouds and produce $\gamma$-rays via decays of $\pi^0$ \citep{2011ApJ...734...28A}.
We therefore checked the archive of the Fermi Large-Area Telescope for a $\gamma$-ray source at the 
position of G308.4--1.4. The Fermi $\gamma$-ray satellite monitors the sky continuously since August 2008. 
However, no source has been detected close to G308.4--1.4 in the Fermi-LAT Second Source Catalog 
\citep{2012ApJS..199...31N} and investigating the images ourselves we found no $\gamma$-ray 
emission matching the X-ray morphology.

\section{Discussion and Conclusion}\label{kap:discus_concl}

Comparing the fitted $n_\mathrm{H}$ with the integrated hydrogen column density towards the source of \citet{2005A&A...440..775K} 
and \citet{1990ARA&A..28..215D} places G308.4--1.4 within our Galaxy.

The plasma model which is used to fit the energy spectrum of G308.4--1.4 was especially designed to improve 
the modeling of X-ray spectra of SNR. Other plasma models assume ionization equilibrium and are therefore not 
able to reproduce the spectra of young SNRs which are not old enough to reach ionization equilibrium 
\citep{2001ApJ...548..820B}. Furthermore, the temperature of the plasma of approximately 6 million Kelvin 
is in the typical range for SNRs \citep{1987soap.conf..226M}.

The analysis of the archival ATCA radio data revealed non-thermal emission with a spectral index of 
$\approx -0.7$, which is typical for young to middle-aged SNRs \citep{1991supe.conf..675D}. The spectral index has a 
large error due to the noise in the 2.5 GHz observation but is still in agreement with the expected value  of 0.4 to 0.5 for 
middle-aged SNR. 

Infrared radiation from SNRs is expected to come either from thermal emission 
of dust grains in the hot plasma which are heated through collision or from the forbidden
lines of elements such as Neon, Oxygen and Iron \citep{2008PASJ...60S.453S}. Several authors could 
show that by comparing the infrared and radio emission it is possible to distinguish between SNRs and HII-regions 
\citep{1987A&AS...71...63F, 1989MNRAS.237..381B, 2011AJ....142...47P}. Moreover, in the work by 
\citet{2011AJ....142...47P} the ratio between the flux density in the MIPSGAL 24-$\mu$m 
\citep{2009PASP..121...76C} and the 1.4-GHz band was used. They find that the ratio is $\leq 10$ for SNR and larger than 
30 for HII-regions. In the case of SNR G308.4--1.4 the 22 $\mu$m band 
was observed which covers roughly the wavelength range from 20 $\mu$m to 27 $\mu$m. 
Thus, the band is comparable with the MIPSGAL 24 $\mu$m band which is sensitive for infrared emission in the range from
20 $\mu$m to 30 $\mu$m \citep{2011AJ....142...47P}. The 22-$\mu$m 
to 1.4-GHz ratio for G308.4--1.4 is $\approx 10$. Hence, the infrared emission matching the radio and 
X-ray morphology and the measured 22-$\mu$m to 1.4-GHz ratio gives further support for G308.4--1.4 being a SNR.

The lack of $\gamma$-ray emission from G308.4--1.4 can be explained by the poor spatial resolution of the
Fermi-LAT telescope ($\sim 1^\circ$ at 1 GeV), which makes it hard to distinguish faint 
extended sources from the diffuse emission of the Galactic plane.
From the 274 known SNRs only 7 have been detected by Fermi-LAT so far \citep{2012ApJS..199...31N}.

There are various physical scenarios to cause a plasma to emit X-rays. However, the two emission 
models for hot diffuse \citep[\sffamily MEKAL\normalfont,][]{1986A&AS...65..511M} and collisionally-ionized diffuse gas 
\citep[\sffamily APEC\normalfont,][]{2001ApJ...556L..91S} which 
are often used to describe the X-ray emission of Galaxy Clusters (ClGs)  cannot 
reproduce the spectrum of G308.4--1.4. Moreover, the temperature of ClGs is normally several $10^7$ K \citep{2010A&ARv..18..127B}, 
one order of magnitude higher than the fitted value. Thus, we can conclude that G308.4--1.4 is not a ClG.
In addition, the possibility of G308.4--1.4 being a planetary nebula (PN) can be ruled out as the X-ray flux 
of G308.4--1.4 is three orders of magnitude higher than the flux of the brightest PN detected in
X-rays, which is at a distance of $\approx 1$ kpc \citep{2006ESASP.604...85G} and the small 22-$\mu$m 
to 1.4-GHz flux ratio. Furthermore, the temperature of G308.4--1.4 is a factor of two higher than the highest measured 
temperature of the hotter, type 2 PNe \citep{2000ApJS..129..295G}.

Based on these results we conclude that G308.4--1.4 is indeed a supernova remnant. Only the eastern part of the remnant 
can be seen in the radio, infrared and X-ray regime which is is not unusual for SNR \citep[e.g.,][]{2011ApJ...734...86C}. A reason 
for that could be that its shock wave is expanding in more dense ISM in the east or the emission in the western part is absorbed by ISM 
which is in front of the SNR. Looking at Fig.\ref{fig:infrared} it can be seen that there is an extended and infrared bright region 
in the north-western part of the SNR whose emission is no correlated with the SNR. 

Using the inferred spectral parameters we can estimate fundamental
characteristics of the remnant, such as the distance from earth and its age.

\subsection{Distance} \label{chap.distance}

To derive a rough estimate for the distance to SNR G308.4--1.4 we use a method described by various authors 
\citep{1999ApJ...511..274S,2001ApJ...561..308S,2003ApJ...591L.143G}: From the distribution of mean color excess 
$\langle E_\mathrm{B-V}\rangle$ per kiloparsec derived by \citet{1978A&A....64..367L} we found 
$\langle E_\mathrm{B-V}\rangle= (0.3\pm 0.1)$ mag kpc$^{-1}$. Using the fitted hydrogen column density 
for the whole remnant $n_\mathrm{H}=(1.02\pm 0.04) \times 10^{22}$ cm$^{-2}$, the relation between the 
hydrogen column density $n_\mathrm{H}$ and the visual extinction $A_\mathrm{V}$ of \citet{1995A&A...293..889P} 
$n_\mathrm{H}=(1.79\pm 0.03)\times 10^{21}$ cm$^{-2} A_\mathrm{V}$  and adopting the relation between 
$A_\mathrm{V}$ and the color excess $A_\mathrm{V}/\langle E_\mathrm{B-V}\rangle=(3.2\pm 0.2) $ 
\citep{2007hsaa.book.....Z} the distance to G308.4--1.4 is $d=5.9\pm 2.0$ kpc. We note, that 
this distance is a lower boundary of the real distance. This is due to the fact that the mean 
color excess $\langle E_\mathrm{B-V}\rangle$ which we used was derived for reddening layer up 
to 2 kpc. In the case of distance estimates for Galactic PNe the reddening inferred distances 
are mostly significantly smaller than distances derived with other methods \citep{2006RMxAA..42..229P}. 
This leads us to assume that $d\approx 6$ kpc is the lower bound limit for the distance to SNR G308.4--1.4. 

Another approach to determine the distance to G308.4--1.4 uses the fact that the radius of the remnant $R_\mathrm{s}$ can be determined in two 
ways, where $R_\mathrm{s}$ depends on the distance $d$ with different exponents. The first equation is a result of the Sedov analysis discussed in the next chapter 
(equation \ref{eq.Rs1}) and depends on $d^{1/6}$. The second equation to determine $d$ is simply the comparison of the radius of the remnant in 
pc ($R_\mathrm{s}$) and in arcmin ($\theta$), $d=R_\mathrm{s}/\theta$, which depends linearly on the distance unit $d$.
Using this dependencies we derive
\begin{equation}
 d=7420 \cdot (\theta~[\mathrm{arcmin}])^{-3/5}\left(\frac{E_{51}}{T_\mathrm{s}[\mathrm{K}]}\right)^{2/5}\left(\frac{f}{\mathrm{Norm}}
\right)^{1/5}=9.8^{+0.9}_{-0.7}~\mathrm{kpc},
\end{equation}
where $T_\mathrm{s}$ is the post-shock temperature and $E_{51}$ the SN explosion released kinetic energy in units 
of $10^{51}$ ergs.

Furthermore, with the help of the ATCA radio observations and the public available HI data from the Southern Galactic Plane Survey 
\citep[SGPS;][]{2005ApJS..158..178M} it is possible to constrain the distance using the Galactic rotation model of 
\citet{1989ApJ...342..272F}. For the region around G308.4--1.4 the SGPS archive provides ATCA data observed with a beam size of 
$130''\times 130''$ and a $1\sigma$ sensitivity of $\sim 1.6$ K. However, the data covers only the northern part of the remnant. 
In the SGPS archive there are also Parkes data covering the region around the SNR, but with an angular resolution of $15'$. 
This beam is twice the size of the remnants diameter. Therefore, we choose to use the ATCA data for the further analysis. 

\begin{figure}
  \resizebox{\hsize}{!}{\includegraphics{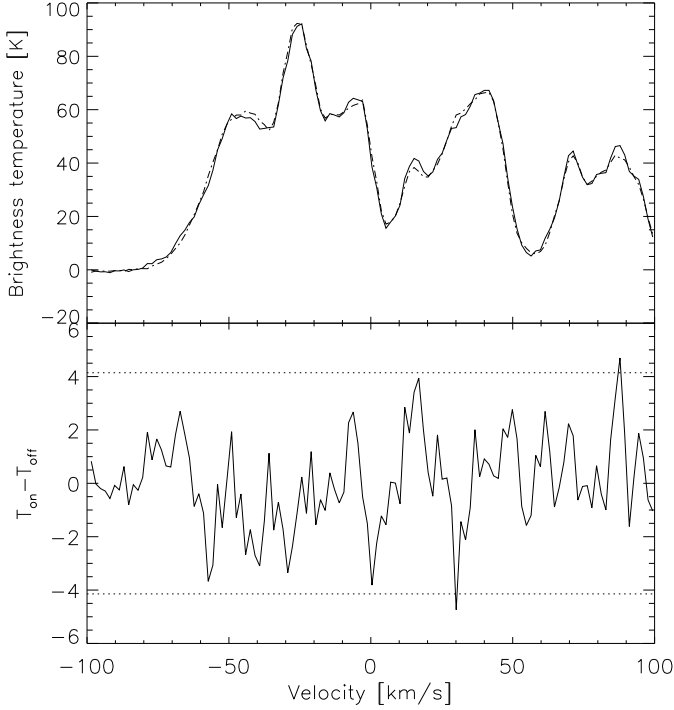}}
  \caption{Top: HI spectrum on- and off-source (continuous and dotted line, respectively) in the direction of SNR G308.4--1.4. 
Bottom: Difference profile with the $3\sigma$ level indicated by a dotted vertical line.}
  \label{fig:hi}
\end{figure}

First, we convolved the radio continuum image at 1.4 GHz to the same beam size as the SGPS data. We assumed that the emission of the 
remnant is not varying significantly across the emission region. Second, the convolved image was used to generate an on-source spectrum 
of SNR G308.4--1.4 by averaging the HI emission for every velocity bin in the HI data cube where the radio continuum emission was above 
5 mJy. The off-source spectrum was computed by taking the average of the emission in the surrounding of the on-source region of the remnant. 
These two profiles as well as the difference plot between these two profiles are shown in Fig. \ref{fig:hi}.
The difference profile is dominated by drifts in the zero level. This is due to the fact that the source is at the border of the 
image and as we are using public data the pointing is not perfect for the method applied. In addition to the drifts we detect one 
absorption feature with a significance of $>3\sigma$ at $V_r\sim 30$ km/s. With the standard IAU parameter for the distance to 
the center of the Galaxy $R_0=8.5$ kpc and the solar orbit velocity of $V_0=220$ km/s derived by \citet{1986MNRAS.221.1023K} and the already 
mentioned Galactic rotation model we deduce a distance to G308.4--1.4 of $2.0\pm0.6$ kpc and $12.5\pm0.7$ kpc. For the error estimate 
we assume a uncertainty in the velocity-to-distance conversion of 7 km/s as suggested by various authors \citep[e.g.,][]{1988ApJ...333..332C} and 
we use the FWHM of the feature as the uncertainty in the velocity determination. Based on the calculated lower limit for the distance 
of the remnant we can rule out the lower value as unreasonable. The low significance of the absorption feature leads to 
the conclusion that the Sedov-analysis based distance approximation is the more reliable estimate for the distance to SNR G308.4--1.4.
Thus, in the following all important quantities are given in units of $d_{9.8}=d/9.8$ kpc.
Using the flux values deduced for G308.4--1.4 we compute its X-ray luminosity to 
be $L_X^{0.5-4}=1.8_{-0.2}^{+0.3}\times 10^{36}~d_{9.8}^{-2}$ erg/s. 

\subsection{Age of SNR G308.4--1.4}

First we had to deduce the post-shock hydrogen density $N_\mathrm{H}$. Using the normalization constant derived in the
model fit of a collisional plasma which is in non-equilibrium (\sffamily VGNEI\normalfont) and assuming that the 
hydrogen and electron densities are constant over 
the volume $V$ of the emission region we derived the following relation: 
$N_\mathrm{e}N_\mathrm{H}=Norm \cdot \frac{\pi D_A^2}{10^{-14} V}$ cm$^{-6}$. $V$ can be approximated by a sphere with radius 
$r\approx \frac{4.2'\pi}{180}\cdot D$ of which only a fraction of $\sim \frac{136}{360}$ is bright enough to 
be detected in the X-ray regime. Therefore, the volume is $V=1.14\times 10^{54}~(r[\mathrm{arcmin}])^3 d^3_{kpc} \mathrm{cm}^3=7.98\times 10^{58} 
d^3_{9.8} \mathrm{cm}^3$. Here $d_{kpc}$ is the distance to G308.4--1.4 in kpc. For a fully ionized 
plasma with cosmic abundances ($\approx 10\%$ He) the ratio between hydrogen and electron density can be estimated 
by $N_\mathrm{H}/N_\mathrm{e}\approx 0.8$. Thus, the post-shock hydrogen density $N_\mathrm{H}=(0.77 \pm 0.11)~d^{-1/2}_{9.8}$ cm$^{-3}$ 
and the corresponding swept-up mass  $M=1.4N_\mathrm{H}m_\mathrm{H}V \sim 71.7 d^{5/2}_{9.8} M_\odot$ \citep{2002ApJ...580..904S}.  

The age of the remnant can be approximated by \citep{1987soap.conf..226M}
\begin{equation}
t=2.71\times 10^9 \left(\frac{E_{51}}{N_\mathrm{H}}\right)^\frac{1}{3} T_\mathrm{s}^{-\frac{6}{5}}d_{9.8}^\frac{1}{6}~\mathrm{yrs}.
\end{equation}
Assuming that the explosion energy is equal to the canonical 
value of $10^{51}$ ergs and the fitted temperature is approximately $T_\mathrm{s}$, the derived value for the hydrogen 
density suggests an age of $\approx 5000 - 7500$ years for SNR G308.4--1.4. Using this result we can calculate the theoretical 
radius of the remnant \citep{1977ASSL...66...29C}  
\begin{equation}\label{eq.Rs1}
R_\mathrm{s} = 0.34 \left(\frac{E_{51}}{N_\mathrm{H}}\right)^\frac{1}{5} t^\frac{2}{5}~\mathrm{pc} = 11.9^{+1.9}_{-1.8} d_{9.8}^\frac{1}{6}~\mathrm{pc}
\end{equation}
and the shock velocity \citep{1959sdmm.book.....S}
\begin{equation}
v_\mathrm{s}=\frac{2}{5} \frac{R_\mathrm{s}}{t} = 730^{+190}_{-160}~\mathrm{km/s},
\end{equation}
which is independent on the assumed distance. The age for the transition from the Sedov-phase 
into the radiative phase is about 29 kyrs \citep{1998ApJ...500..342B}. The fact that the age of G308.4--1.4 is 
much lower than the transition time scale suggest that G308.4--1.4 is in the Sedov-phase. Furthermore, the energy 
spectrum of the remnant is in agreement with G308.4--1.4 being in the Sedov-phase.

However, the derived parameters should only be seen as rough estimates as we could not determine whether
a full temperature equilibrium is already established. If not, a generally higher ion temperature which 
determines the SNR dynamics would lead to a higher shock velocity and subsequently to an e.g.~overestimation of the remnant's age.  

\subsection{The central sources CXOU J134124.22--634352.0 and CXOU J134127.12--634327.7}\label{kap.dis_point_src}

No radio counterpart for CXOU J134124.22--634352.0 has been detected in the ATCA and SUMSS observations 
of G308.4--1.4. An explanation could be that the variable source is also 
variable in the radio band and therefore the source was missed in all previous observations of this region. To decide 
if this is the case new observations are needed. 

Additionally, CXOU J134124.22--634352.0 has no compact/extended X-ray emission as seen in other young and powerful pulsars. Its best fit 
spectral model consists of two blackbodies. Thus, source \# 1 shows some properties 
which have been also seen in central compact objects \citep[CCO, see][for a review]{2008AIPC..983..320G}. 
CCOs are young neutron stars which are located in SNRs having an age of less than 10 kyrs. They have X-ray luminosities
between $10^{32}$ and $10^{33}$ erg/s in the energy range 0.5-10 keV. No CCO has a detected counterpart in the 
radio and optical band so far. For one member of the CCO group, 1E 161348--5055 in RCW 103, flares have been detected 
in which the flux varies between 0.8 and $60 \times 10^{-12}$ ergs cm$^{-2}$ s$^{-1}$. This is 
comparable to source \# 1 where the maximal flux is $\approx 10^{-11}$ ergs cm$^{-2}$ s$^{-1}$ 
and the upper limit derived from the SWIFT observations is $\approx 5 \times 10^{-13}$ ergs cm$^{-2}$ s$^{-1}$.

Using the distance derived in chapter \ref{chap.distance} the luminosity of CXOU J134124.22--634352.0 is
$L_X^{0.5-10}=2.5_{-0.6}^{+1.2}\times 10^{34}~d_{9.8}^{-2}$ erg/s. This is slightly higher than the derived luminosity 
for the normal population of CCO \citep[p.121]{2009Becker}, only the flaring CCO in RCW 103 has a comparable luminosity of 
$0.1-8 \times 10^{34}$ erg/s \citep{2008AIPC..983..311D}.

In the ROSAT All-Sky Survey \citet{1999Krautter} found for all X-ray bright stars 
with an optical counterpart a ratio of $\log(f_X/f_V)=-2.46 \pm 1.27$. 
Therefore, the X-ray-to-visual flux ratio of 1.55 and the fitted $n_\mathrm{H}$, which is in 
the order of what has been observed for G308.4--1.4, suggest a CCO interpretation
for source \#~1.

Assuming that the source is indeed the compact remnant we can derive the proper motion of the object 
$\mu=\sqrt{(RA_c-RA_{\# 1})^2+(DEC_c-DEC_{\# 1})^2}/t = (45\pm 4)''/t$ using the inferred center and the age of the SNR. We 
deduce the proper motion to be  $\mu=7 \pm 2$ mas/yr $= 320 \pm 100$ km/s. The upper limit on $\mu$ is higher than 
the mean two-dimensional proper motion of pulsars in SNRs of $\sim 227$ km/s \citep{2005MNRAS.360..974H}, but 
would not be an exception. For example, the CCO RX J0822-4300 in the SNR Puppis A has a 2d proper 
motion of $672 \pm 115$ km/s \citep{2012arXiv1204.3510B}.

Nevertheless, the lower limit of the inferred emitting radius of the blackbody in the double blackbody or the 
power law-blackbody spectral fit is in the order of the expected value for the neutron star radius of 10 to 20 km only 
if the source is at a distance larger than 14.5 kpc (see Fig. \ref{fig.R_vs_d}). Moreover, we found 
an infrared and optical source which position is consistent with source \# 1. The spectral distribution of the 
source allows us to fit a collisionally-ionized diffuse gas model (\sffamily APEC\normalfont) which 
\citet{2005ApJ...621..398O} used to model the energy distribution in the X-ray band of the M dwarf flare star 
EV Lacertae. In addition, the fitted hydrogen column density is consistent to that of a nearby star and the X-ray-to-visual flux ratio
is 0.25, only $2\sigma$ higher than the value for stars derived by \citet{1999Krautter}.

Using the stellar density in direction of G308.4--1.4 we can derive the chance association of the observed 2MASS source with 
the compact remnant candidate. The chance association is given by 
$P_{coin}=\frac{N}{l_{RA} l_{Decl.}} \pi~ \delta RA~ \delta DEC$, where $N$ is the number of sources detected within 
a rectangular region of length $l_{RA}$ and width $l_{Decl.}$ and $\delta RA$, $\delta DEC$ are the errors in the position 
of the source. The 2MASS catalog contains 2206 point-like sources 
in a box with side length of 5 arcmin around the center of the remnant and the chance association is 1.3\%.
Thus, the possibility of a false association between the X-ray and the 2MASS source cannot be excluded. 

Until now, we did not discuss the western radio arc which has no counterpart in the X-ray and infrared image. The arc could 
be interpreted as a relativistic radio jet of a source at the geometrical center of the SNR. If so, the central source 
is either a black hole of stellar mass or a neutron star in a binary system \citep[~and reference therein]{1999ARA&A..37..409M}. 
Especially the possibility that the central source is a black hole is an interesting speculation which, if proven, would 
mean that G308.4--1.4 is the remnant of a Type II core-collapse SN. Thus, would be the first one known which left over a black hole.


To obtain a rough estimate of the source flux of the other source in the center of the remnant, source \# 10, we assumed that the 
source is a CCO, should have at least a spectrum with a blackbody of temperature 2.6 million Kelvin as the CCO in 
Puppis-A \citep[cf. Table 6.4,][]{2009Becker}. Using the WebPIMMS 
tool with the source count rate and the fitted $n_\mathrm{H}$ of the SNR 
the flux in the $0.3-3.5$ keV range is $f_X \approx 2.2\times 10^{-14}$ ergs cm$^{-2}$s$^{-1}$ and 
in the $0.5-10$ keV range is $f_X \approx 1.9\times 10^{-14}$ ergs cm$^{-2}$s$^{-1}$. The normalization 
of a blackbody with this flux, temperature and $n_\mathrm{H}$ is $2.5\times 10^{-5}$ which corresponds to 
an emitting radius of the source of $\approx 9 d_{9.8}^{-1}$ km. This value is in perfect agreement with the expected 
value for the neutron star radius. The corresponding luminosity in the energy band $0.5-10$ keV is 
$L_X^{0.5-10}=2.1\times 10^{32}~d_{9.8}^{-2}$ erg/s, a typical value for CCOs. Furthermore, the X-ray-to-visual 
flux ratio is $\log(f_X/f_V)>0.11$ taking into account that no optical source was detected in the USNO-B1.0 
catalog and the limiting magnitude of this catalog is 21 \citep{2003AJ....125..984M}.
Thus, no clear evidence for the origin of the emission of source \# 10 could be found.

Finally we mention that the detector support structure of Chandra covers the
central part of G308.4--1.4 (see Fig. \ref{fig.chandra_rgb}). It is therefore not
excluded from our observations that other faint X-ray point sources are located 
at this position. However, the non-detection of a source in the SWIFT data 
at the assumed position of the remnant's expansions center sets a strict upper limit of 
$5 \times 10^{-13}$ ergs cm$^{-2}$ s$^{-1}$ on the flux of such an object.

\section{Summary}\label{kap:Summary}

The high resolution imaging and spectroscopic capability of Chandra revealed that the X-ray emission of 
the source is an extended plasma and G308.4--1.4 is indeed a supernova remnant. The emission region is spherically 
symmetric with a diameter of 8.4 arc-minutes and an excess emission towards the east. 

The radio morphology at 0.843, 1.384 and 2.496 GHz is characterized by two arcs. The eastern arc matches well with the X-ray and 
infrared contours. The second arc which is bending in the west and decreasing brightness towards the center of G308.4--1.4 has no 
counterpart in other wavelengths. The remnant without the two bright knots has a spectral index of $\alpha=-0.7\pm0.2$

\begin{table}
\renewcommand{\arraystretch}{1.25}
\begin{footnotesize}
\caption{Fundamental parameters of SNR G308.4--1.4.}
\label{tab:imp_para}
\begin{tabular}{cc}
\hline\hline
SNR & \\ \hline
distance & $9.8^{+0.9}_{-0.7}$ kpc\\
age & $5000-7500$ yrs\\
radius & $\sim 4.2'$/ $11.9^{+1.9}_{-1.8} d_{9.8}^{1/6}$ pc \\
expansion velocity & $730^{+190}_{-160}$ km/s\\
\\[-2ex]\hline\\[-1ex]
\end{tabular}
\end{footnotesize}
\end{table}

The Sedov analysis lead to the conclusion that the SNR is about 6200 years old and is expanding with a velocity in the order 
of $\approx 730$ km/s. All fundamental parameters of the SNR are summarized in Table \ref{tab:imp_para}. 

The X-ray point-like source CXOU J134124.22--634352.0, located close to the geometrical center of the SNR
G308.4--1.4, is seen to exhibit variable and flaring X-ray emission. However, no certain conclusion on the origin of this 
source and the other central source CXOU J134127.12--634327.7 could be drawn. Deeper observations of the central part of SNR G308.4--1.4 are 
needed to resolve the obvious ambiguities.

\begin{acknowledgements} 
 T.P. acknowledges support from and participation in the International Max-Planck 
Research School on Astrophysics at the Ludwig-Maximilians University. The authors acknowledge 
discussions with M.~D. {Filipovi{\'c}}.
\end{acknowledgements}

\bibliographystyle{aa} 
\bibliography{literatur} 

\end{document}